\documentclass{article}
\usepackage[margin=1.55in]{geometry} 
\usepackage{pslatex} 
\usepackage{graphicx}
\usepackage{authblk}
\usepackage[caption=false]{subfig}
\usepackage{hyperref}
\usepackage[utf8]{inputenc}
\usepackage{amsfonts}
\usepackage{booktabs}
\usepackage{siunitx}
\usepackage{multirow} 
\usepackage[normalem]{ulem}
\useunder{\uline}{\ul}{}
\usepackage[table,xcdraw]{xcolor}
\begin{document}
\title{Modeling tax distribution in metropolitan regions with PolicySpace}
\author[1]{Bernardo Alves Furtado}
\affil[1]{Institute for Applied Economic Research (IPEA), Brazil}
\affil[1]{National Council of Research (CNPq)}
\maketitle              
\abstract{Brazilian executive body has consistently vetoed legislative initiatives easing creation and emancipation of municipalities. The literature lists evidence of the negative results of municipal fragmentation, especially so for metropolitan regions. In order to provide evidences for the argument of metropolitan union, this paper  quantifies the quality of life of metropolitan citizens in the face of four alternative rules of distribution of municipal tax collection. Methodologically, a validated agent-based spatial model is simulated. On top of that, econometric models are tested using real exogenous variables and simulated data. Results suggest two central conclusions. First, the progressiveness of the Municipal Participation Fund and its relevance to a better quality of life in metropolitan municipalities is confirmed. Second, municipal financial merging would improve citizens' quality of life, compared to the \textit{status quo} for 23 Brazilian metropolises. Further, the paper presents quantitative evidence that allows comparing alternative tax distributions for each of the 40 simulated metropolises, identifying more efficient forms of fiscal distribution and contributing to the literature and to contemporary parliamentary debate.

}

\section{Introduction}
The post-Constitution period of 1988 was fruitful for the dismemberment and creation of new municipalities in Brazil. In fact, the number of municipalities in Brazil increased by about 35\%, with an increase of 1,438 new municipalities up to the 2000 Census \cite{fernandes_criacao_2017}. Most of the new municipalities are small, with a population of less than 10 thousand inhabitants. Municipalities with a reduced population have less capacity to offer public services and collect local taxes \cite{marenco_capacidade_2017}.

A break on new municipalities creation was achieved only with the publication of the Constitutional Amendment number 15 of 1996\footnote{\url{http://www.planalto.gov.br/ccivil_03/Constituicao/Emendas/Emc/emc15.htm}}, which establishes stricter rules. A recent veto message summarizes the arguments by which the creation and dismemberment of municipalities impacts the municipal Public Administration as a whole: (a) continued growth of expenses and administrative structure, (b) maintenance of revenues and (c) pulverization of resources transferred from the Union to the municipalities.

From the point of view of each municipality independently, the dismemberment can be positive since (a) a new administrative structure is created, opening new positions in the Executive and Legislative (City Hall), and (b) the pulverization of resources is made with the gain of the 'new' municipalities, to the detriment of the rest of the state municipalities, given the distribution rules of the Municipal Participation Fund (MPF), given the maintenance of the amount transferred by the Union.

This debate between possible advantages of dismemberment and its costs is already present in the literature and is discussed in the next section. However, we have no knowledge of quantitative analyzes that simulate results from alternative resource distributions between municipalities. Thus, the proposal of this paper is to present simulations of alternative tax distribution for municipalities of Brazilian metropolitan regions. Methodologically, the simulation is done by a validated agent-based model \cite{furtado_policyspace:_2018}. Results indicate that, for the vast majority of the analyzed cases, the merger of the municipalities of the metropolitan regions is beneficial, albeit of small magnitude, in terms of gains in quality of life for its inhabitants in the period of 20 years. The exercise also reinforces the relevance of the progressive effects of MPF in the current configuration.

The contribution of this text lies in the presentation of quantitative evidence - based on empirical data - that incorporates dynamic effects of redistribution of the amount of public resources in alternative ways. With the results, we confirm other disciplinary aspects of the literature that argue that in most cases the dismemberment is not socially beneficial.

Besides the introduction, we briefly discuss the literature (\ref{context}), describe the methodology used (\ref{methodology}), and present the results (\ref{results}). We conclude with some final considerations (\ref{final}).

\section{Context and Literature}\label{context}
This section summarizes the recent debate in the legislature on emancipationist proposals and reports favorable and opposing arguments. The research proposal that derives from the context analysis and the availability of methodologies concludes the section.

The theme of creation and municipal emancipation remains in evidence more than twenty years after the publication of Constitutional Amendment 15 of 1996. Two more attempts were made after veto 550 of 2013. Bill 104 in the Senate (and Draft Complementary Law 397 of 2014) was again vetoed in full in message 250 of August 26, 2014. The veto states that "fiscal responsibility" and the imbalance of possible redistributions as its reasons \footnote{\url{http://www.planalto.gov.br/CCIVIL_03/_Ato2011-2014/2014/Msg/Vet/VET-250.htm}}. In another parliamentary attempt, Senate Bill 199 of 2015 was approved in plenary and it is in process of urgency as Draft Complementary Law 137/2015 of the Chamber of Deputies having received six other initiative attachments. \footnote{\url{http://www.camara.gov.br/proposicoesWeb/fichadetramitacao?idProposicao=1594899}}.

A recent report currently in discussion in the House suggests that the creation of municipalities can take place between the possession of the elected mayor and the last day of the year before the year of municipal elections. Excluding, therefore, only the years in which there is municipal election. Further, the report proposes some criteria for the approval of new municipalities: (a) minimum population quota for the new municipalities and for the remaining municipalities (b) that the dismemberment is supported by Municipal Feasibility Study (EVM), and that the EVM should include economic and financial analysis, among other requirements.

Fundamentally, the representatives propose similar criteria to those who were in effect until 1996 with little more rigorous enforcement and population minimums. Most likely, the approval of Bill 199 would generate hundreds of new municipalities \cite{sachsida_criacao_2013}. This indication of creation of new municipalities seems counterproductive to the debate in the literature.

In the context of metropolitan areas, the Organization for Economic Co-operation and Development (OECD) defines employment basins as "functional regions" \cite{ahrend_what_2014}. 
They refer to economically unique regions, distributed throughout the diffuse territory in different political-administrative contexts. In Brazil these functional regions are called Areas of Population Concentration (APCs) by IBGE, the official statistics bureau \cite{ibge._ministerio_do_planejamento_arranjos_2015}. 

APCs fragmentation among several municipalities bring various negative aspects. First, more fragmented metropolis are less productivity \cite{ahrend_what_2014}. Further, the peripheral cities experience increased levels of violence whereas receiving lower quality of public services such as transport, or housing. From a strictly fiscal point of view, the literature suggests not only that metropolitan municipalities receive more per capita tax revenue \cite{furtado_fatos_2013}, but also that they are more efficient in their application. Fernandes describes metropolitan municipalities weak financial capacity.  

\begin{quotation}
One of the major problems faced by the Brazilian federation is the municipal autonomy due to the low fiscal, financial and institutional capacity of local governments to absorb and account for all their constitutional functions, including urban policy \cite[p.3]{fernandes_criacao_2017}.
\end{quotation}

The literature suggests that that fragmented metropolitan regions with multiple municipalities: (a) are less productive; (b) present polarized capacity between the metropolis and its neighbors; (c) concentrate diseconomies of agglomeration in the peripheries (violence and congestion); (d) concentrate agglomeration economies at the main city and (e) restrict scale gains in the provision of network services, such as transportation, basic sanitation (water and sewage) and land use and occupation (social housing).

However, there are spatially local gains generated with emancipation. The arguments suggest that the territorial extension of the municipality may make it difficult to serve distant districts with public services. Economically, there is favorable redistribution of public resources to the new municipalities and also a whole new political-administrative structure, which generally leads to actual improvement of the public service. In basic sanitation, for example, Rocha and colleagues \cite{rocha_descentralizacao_2017} indicate a 12\% increase in investments in the sector for new municipalities. Wanderley \cite{wanderley_emancipacoes_2008} indicates gains in health and education indicators, while not detecting a significant loss in the remaining municipalities.

Hence, we argue for the need of a more informed analysis about the gains and losses for the whole of society in alternative redistributions of tax resources. One way to operationalize economic analysis as open dynamic systems is through agent-based modeling (ABM). ABM is characterized by the computational environment replication of agents characterized by attributes that evolve their characteristics according to deterministic and explicit rules. 

In this paper, we use an agent-based model to dynamically evaluate relevant markets on which municipal taxes are collected and distributed as public services. The model intends to replicate in time and space the patterns and the financial environment in which alternative distributions of taxes can be simulated, if not \textit{in vitro}, at least \textit{in silica}.

Our proposal can be summarized as follows:
\begin{enumerate}
    \item Build a spatially empirical environment which simulates three markets upon which five distinct taxes are applied to for the years 2000-2020. Thus replicating the \textit{status quo} of generation and distribution of tax revenues for 40 Brazilian metropolitan regions.
    \item Validate the results so that the model becomes able to answer the public policy research question. Specifically: are there other forms of distribution of tax resources that improve the quality of life of the inhabitants at the metropolitan level?
    \item Simulate alternatives and compare with current situation.
    \item Present and discuss the results, in light of the available literature.
\end{enumerate}

\section{Methodology: agent-based modeling}\label{methodology}

An agent-based model is the implementation of an artificial computing environment in which agents interact in time and space. Or as described by Epstein and Axtell \cite{epstein_growing_1996}: ABM is a discrete and dynamic temporal system, explained by simultaneous generic equations.

Among the advantages of using ABMs is the low cost (\textit{in silico} experimentation), the possibility of conducting experiments (what-if questions), and its explicit spatial, dynamic and modular construction which allows other users to develop additional resources to platform available. In addition, open source software allows the reproducibility of the results and the full comprehension of the mechanisms used in the model.

Dawid and Delli Gatti \cite{dawid_agent-based_2018} propose seven major families of agent-based modeling and its applications in public policy and economics. PolicySpace \cite{furtado_policyspace:_2018} is a proposal of economic-spatial modeling that fits the Lengnick \cite{lengnick_agent-based_2013,gaffeo_adaptive_2008} family. However, PolicySpace \footnote{PolicySpace is open source, available at \url{Github.com/BAFurtado/policyspace}.} differs from previous contributions as it adds intraurban spatial analysis, household mobility, population dynamics and the use of distance as a criterion of choice in consumer and labor markets.

We use PolicySpace to systematically analyze four tax revenue distribution alternatives between municipalities 40 Brazilian metropolitan regions. After confirming the validity of the model, econometric tests are applied, using real data and simulated data to evaluate the strength of the results.

\subsection{PolicySpace in a nutshell}
PolicySpace counts on agents - citizens - who offer themselves in the labor market and organize into families. Families participate in the consumer and real estate market, so they are mobile and may change the location of their home. Firms employ workers and offer a homogeneous product in the consumer market. They compete by prices in the consumer market and for more skilled workers in the labor market. Municipal governments are constituted according to geospatial real data of the IBGE. Municipalities invest the taxes collected and received by transfers in the improvement of the quality of life of its citizens in a linear way, weighted by current population. The model runs every month from 2000 to 2020 and the sequence of events happens as follows:

\begin{itemize}
    \item Firms performs its production function based on the number of current employees and their qualification \cite{lengnick_agent-based_2013}.
    \item Population dynamics is executed. Citizens age, die and are born. 
    \item Families save a variable percentage of their income and consume the rest from a sample of firms.
    \item Firms decide on the salaries of their employees \cite{neugart_agent-based_2012}, on prices \cite{seppecher_what_2017}, and on the need to participate in the labor market.
    \item Firms paying higher salaries choose employees first, opting for those most qualified from the sample of candidates. Optionally, a percentage of the workforce can be chosen by criterion of proximity of the residence of the candidate to the employing firm.
    \item Monthly, some families enter as buyers in the real estate market \cite{jordan_agent-based_2012} and there are always more empty houses than occupied ones \cite{nadalin_concentracao_2018}.
    \item The model uses hedonic housing pricing \cite{rosen_hedonic_1974}. Final transaction price is the average between the calculated price of the residence and the buying family offer.
    \item Taxes are collected: (i) at the moment of consumption, (ii) on wages paid (iii) on company profits, (iv) on property, and (v) on real estate transactions. 
\end{itemize}
\item 

Validation of PolicySpace is proposed in three successive steps that indicate the robustness of the procedures. The first is the adequacy of macroeconomic indicators. PolicySpace is compatible with large macroeconomic indicators. 

In addition, validated models must be robust to modifications on parameters \cite{galan_errors_2009}. PolicySpace showed no structurally different responses to sensitivity analysis. For example, the test on the increase of the productivity parameter, showed improvement on consumption and the standard of living of families. Further, increase in the number of active families in the real estate market led to the dynamization of the economy as a whole. Finally, it is worth mentioning that rules tests were also carried out. That is, the presence or absence of certain choices of the model were testes and shown to be robust. 

Perhaps, in the case of the research question in this text, the most relevant validating factor is the ability of the simulation to replicate the tax collection results, both in terms of percent of GDP and percentages of each tax in total. Two sources were used for the basis of the comparison. One is based on the literature \cite{afonso_avaliacao_2013} and another on data provided by the National Treasury Department of the Ministry of Finance (STN).

\subsection{Testing strategy for tax alternatives}
We tested four alternative distributions of fiscal resources (\ref{table1}). For each case, PolicySpace is simulated for the full 20-year period, several times and the median results are computed.

\textbf{Case 1} serves as baseline and represents the \textit{status quo}. Consumption tax is distributed at 18.75\% for the municipality of origin and the remainder (81.75\%) is passed on to the state and (theoretically) evenly distributed among the municipalities of the metropolitan region. In effect, there are three criteria of division of taxes: (1) locally - resources collected in the municipality are invested in the own municipality; (2) equally - resources collected in the metropolitan region (APC) are also distributed  equally between the municipalities of the APC, weighted by the population, and (3) MPF - whose proportionality follows the empirically-observed distribution of real MPF. In the MPF rule smaller municipalities receive proportionately more resources.

\begin{table}[h]
\centering
\caption{Alternative tax distributions tested}
\label{table1}
\begin{tabular}{@{}lccccccc@{}}
\toprule
 & \multicolumn{3}{c}{Case 1} & \multicolumn{2}{c}{Case 2} & Case 3 & Case 4 \\ \midrule
Taxes & Municipalities & State & MPF & State & MPF & Municipalities & State \\ \cmidrule(r){1-8} 
Consumption & \cellcolor[HTML]{BFBFBF}0.1875 & \cellcolor[HTML]{BFBFBF}0.8125 & \cellcolor[HTML]{BFBFBF} & 1 &  & \cellcolor[HTML]{BFBFBF}1 & 1 \\
Personal income & \cellcolor[HTML]{BFBFBF} & \cellcolor[HTML]{BFBFBF}0.765 & \cellcolor[HTML]{BFBFBF}0.235 & 0.765 & 0.235 & \cellcolor[HTML]{BFBFBF}1 & 1 \\
Transmission & \cellcolor[HTML]{BFBFBF}1 & \cellcolor[HTML]{BFBFBF} & \cellcolor[HTML]{BFBFBF} & 1 &  & \cellcolor[HTML]{BFBFBF}1 & 1 \\
Company & \cellcolor[HTML]{BFBFBF} & \cellcolor[HTML]{BFBFBF}0.765 & \cellcolor[HTML]{BFBFBF}0.235 & 0.765 & 0.235 & \cellcolor[HTML]{BFBFBF}1 & 1 \\
Property & \cellcolor[HTML]{BFBFBF}1 & \cellcolor[HTML]{BFBFBF} & \cellcolor[HTML]{BFBFBF} & 1 &  & \cellcolor[HTML]{BFBFBF}1 & 1 \\
\cmidrule(r){1-8} 
Criteria & \cellcolor[HTML]{BFBFBF}Locally & \cellcolor[HTML]{BFBFBF}Equally & \cellcolor[HTML]{BFBFBF}MPF & Equally & MPF & \cellcolor[HTML]{BFBFBF}Locally & Equally \\ \bottomrule
\end{tabular}
\end{table}
Source: Adapted from Furtado \cite{furtado_policyspace:_2018}.

In \textbf{Case 2}, APC municipalities function as a single municipality for tax distribution purposes and there are no locally reverted resources. In this case, taxes on consumption, property and property transmission are distributed equally among all the municipalities of the APC, weighted by the population. Taxes on labor and company profits are distributed equally between the municipalities (76.5\%) and the rest is distributed according to the MPF rules for municipalities in the metropolitan area.

Then, two extreme cases are tested in order to enrich the analysis. In \textbf{Case 3}, all taxes are distributed according to the municipalities of origin. That is, there is no progressive redistribution by MPF, nor consumption tax is distributed throughout the municipalities, as if the State was forced to apply all the resources locally.

Finally, in \textbf{Case 4}, 100\% of the amount collected with the consumption tax is distributed equally among the municipalities, weighted by the population.

In short, the \textit{status quo}, \textbf{Case 1}, is compared to the absence of the redistributive MPF (\textbf{Cases 3, 4}) and the fiscal union of municipalities (\textbf{Cases 2, 4}). \textbf{Case 3} is the extreme case wherein all taxes are locally distributed.

The research hypothesis is that \textbf{Case 2} would be the most beneficial for the population with municipal merger and MPF progressiveness. \textbf{Case 3} would be the least beneficial, with all taxes withheld and distributed locally. \textit{Ex ante}, it is not possible to determine whether \textbf{Case 2}, which includes the progressivity of the MPF associated with local taxes would be better or worse than \textbf{Case 4}, which does not include the progressiveness of the MPF, but distributes the taxes equally.

It should be emphasized that the exercise is only a redistribution of funds raised. There is no tax gain due to a supposed reduction of administrative bureaucracy or gains with a supposed improvement in efficiency \cite{gasparini_transferencias_2011}.

\subsection{Econometric strategy}
The econometric strategy seeks to verify if the two parameters of the model: Alternative0 and MPF distribution that compose the four fiscal distribution alternatives (\ref{table1}) are robust as determinants of the quality of life obtained in the simulation. As such, five econometric models are explored, three with simulation data (\textbf{Simul1-3}) and two with real data, exogenous to the model (\textbf{Real1-2}).

In all cases, the dependent variable is the quality of life observed at the end of the simulation period. Also, in all cases, the two parameters Alternative0 and MPF represent the distribution alternatives exactly as depicted in \ref{table1}. Alternative0 is \textbf{true} for the \textit{status quo} and \textbf{false} when the municipalities are together for distribution purposes. MPF is \textbf{true} when the distribution rule is present according to the presence of MPF and \textbf{false} when the rule is not applied. Additionally, dummies of all APCs are used for the \textbf{Simul1-2} cases, with ztextbf{Simul3} testing their absence. In \textbf{Real1-2} cases, information was collected only for the main metropolises \footnote{The complete results (with APCs dummies), the database and the code used are available in \url{https://www.dropbox.com/sh/udke6c196stjzy9/AACljy9Cbb-zmR_-1AQoyFcMa?dl=0}.}. 

In \textbf{Simul2-3} cases, the controls are the variables of average number of workers per firm, firm profit, GDP index value, inflation, unemployment and number of municipalities in the APC. The latter was also used in \textbf{Real2} model. Finally, in \textbf{Real2} model, a HHI index, the log of the population in the APC and the percentage of inhabitants with complete higher education were used.

We highlight that quality of life in the simulation is basically given by the tax collection. Thus, no variables associated with taxes were included in the models. However, it is known that GDP and population are highly correlated with tax collection. GDP (endogenous to simulation) was included in \textbf{Simul2-3} models and the population in \textbf{Real2} model. Nevertheless, \textbf{Simul1} and \textbf{Real1} models only rely on the presence of the two distribution parameters and the dummies of the APCs.

\section{Results} \label{results}
The results are obtained for each APC independently and the \textbf{policy recommendation} can also be different for each case. Indeed, several factors influence the configuration of each APC, including: population, their age cohorts and distribution among the municipalities, the concentration of firms and the qualification of workers, besides the absolute number of municipalities.

We present a systemic result for the set of metropolitan regions (APCs) in Figure 1. In this case, the Quality of Life Index (QLI) values for the last month of the simulation are normalized in order to distinguish more clearly the differences between the results and allow for comparison among APCs. 

\begin{figure}[ht]
\centering
\includegraphics[width=\textwidth]{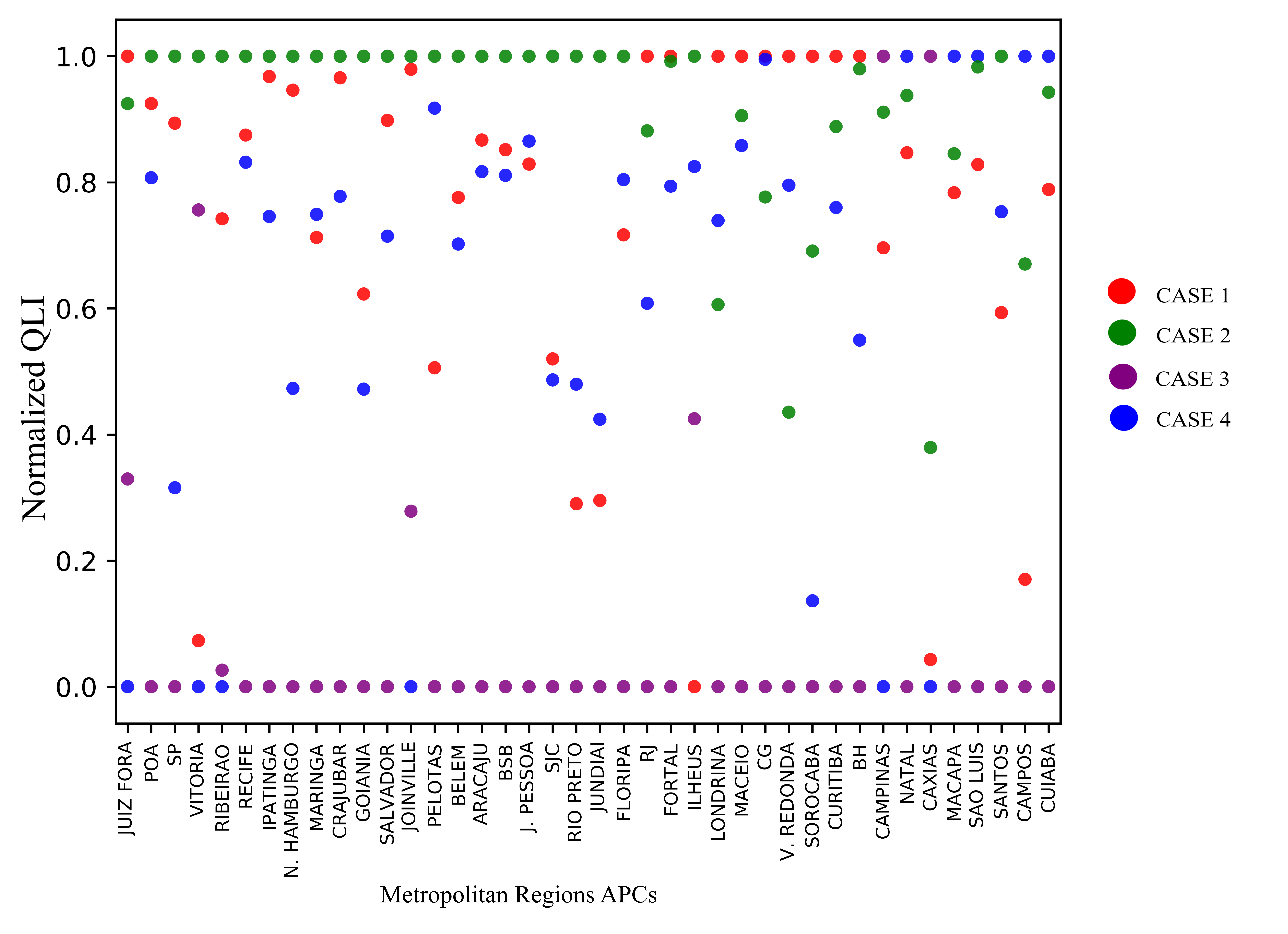}
\caption{Results of the simulation of alternative tax distribution among metropolitan regions (APCs). \textbf{Case 2} which represents the maintenance of the MPF distribution rules and the union of all municipalities in the APC, presents the highest number of maximum values among the options for the metropolitan regions. \textbf{Case 3} which represents the withdrawal of MPF rule leads to worse quality of life results in the greatest number of metropolitan regions. The results refer to simulation runs of 2\% of the population and 3 simulations per APC. Authors' elaboration.}
\end{figure}

Indeed, for 23 APCs, \textbf{Case 2} with municipal merger and MPF maintenance is the result with a better Quality of Life indicator. The current situation (\textbf{Case 1}) shows the best result in 10 APCs\footnote{Note The abbreviations of APCs are: POA: Porto Alegre, N. Hamburg New Hamburg, Crajubar, Crato, Juazeiro and Barbalha, Brazil: São Paulo, BSB: Brasilia, J. Person: João Pessoa, SJC: São José dos Campos, Floripa: Florianópolis, CG: Campina Grande, V. Round: Volta Redonda, Belo Horizonte and Campos: Campos dos Goitacazes.}. Nevertheless, \textbf{Case 3}, in which all resources are distributed locally, has only two APCs with the best QLI results: Campinas and Caxias do Sul. Finally, \textbf{Case 4}, in which the municipalities are considered together, but there is no distribution according to MPF, is better for the case of five APCs.

The results of the econometric exercise suggest that the fiscal distribution alternatives among the municipalities of the studied metropolitan regions are consistent and robust in all models tested (see \ref{table2}). As mentioned earlier, variables \textbf{Alternative0} and \textbf{MPF\_distribution} are the ones being tested. Together they configure the four cases of distribution as described in Table 1\footnote{Note, however, as stated above, that for \textbf{Case 3}, chosen as the extreme case, consumption tax is distributed locally in full.}.

\begin{table}[ht]
\centering
\resizebox{\textwidth}{!}{%
\begin{tabular}{@{}lcccccc@{}}
\toprule
& Simul 1 & Simul 2 & Simul 3 & Real 1 & Real 2 &  \\ \midrule
ALTERNATIVE0 {[}True{]} & -0.01*** & -0.01*** & -0.01* & -0.01*** & -0.01*** \\
 & (0.00) & (0.00) & (0.01) & (0.00) & (0.00) \\
MPF\_DISTRIBUTION {[}True{]} & 0.02*** & 0.01*** & 0.02** & 0.02*** & 0.02*** \\
 & (0.00) & (0.00) & (0.01) & (0.00) & (0.00) \\
Intercept & 0.60*** & 0.9 & 0.73*** & 0.61*** & 0.02*** \\
 & (0.01) & (0.64) & (0.03) & (0.01) & (0.00) \\
area.APC &  &  &  &  & -0.00*** \\
 &  &  &  &  & (0.00) \\
pib.index &  & 0.00** & 0.00 &  &  \\
 &  & (0.00) & (0.00) &  &  \\
inflation &  & -1.55 & 11.31*** &  &  \\
 &  & (1.10) & (3.15) &  &  \\
ln.population.APC &  &  &  &  & 0.01*** \\
 &  &  &  &  & (0.00) \\
number.municipality.APC &  & -0.01 & -0.00 &  & 0.00** \\
 &  & (0.01) & (0.00) &  & (0.00) \\
Note\footnote{Full table available at \url{https://www.dropbox.com/sh/udke6c196stjzy9/AACljy9Cbb-zmR_-1AQoyFcMa?dl=0}} & ... & ... & ... & ... &  \\
\midrule
Log-likelihood & 506.85 & 510.85 & 267.02 & 260.65 & 260.65 \\
R-squared Adj & 0.98 & 0.98 & 0.60 & 0.98 & 0.98 \\
AIC & -931.69 & -929.7 & -516.04 & -477.31 & -477.31 \\
BIC & -806.65 & -789.41 & -488.59 & -424.9 & -424.9 \\
No. Observations & 156 & 156 & 156 & 80 & 80 \\ \bottomrule
\end{tabular}%
}
\caption{Compact results of the econometric tests of the tax distribution alternatives between APCs. Authors' elaboration}
\label{table2}
\end{table}

The interpretation, as expected, indicates that when \textbf{Alternative0} is \textbf{True} - in cases where the municipal division remains as observed - there is loss of Quality of Life (negative sign). This indicates that there is a clear gain for all models when \textbf{Alternative0} option is \textbf{False} and therefore the municipalities are all together for fiscal distribution purposes.

Additionally, the presence of the MPF distribution rule (\textbf{True}) is also beneficial in all cases. It is equivalent to say that when MPF is not present in the fiscal distribution (\textbf{False}), there would be a deterioration of Quality of Life.

The controls of the models lead to different values for the intercept and for the significance of each of the controls. However, they do not change the coefficients or the significance of the distribution rules and present similar adjustments.

\textbf{Simul 2} model, with dummies for each APC and other control variables, appears as the most adjusted among the models simulated considering the log-likelihood criterion. However, \textbf{Simul 1} model, with the dummies only, presents a very similar adjustment and seems more robust, from the point of view of the analysis of the APCs (see full results at \url{https://www.dropbox.com/sh/udke6c196stjzy9/AACljy9Cbb-zmR_-1AQoyFcMa?dl=0}).

Finally, in addition to the descriptive analysis of Figure 1 and Table 2, the simple econometric exercise seems to reinforce the suggestion of this work on the relevance of the presence of MPF and the gains in distributive efficiency of the merger of metropolitan municipalities.

\section{Final considerations}\label{final}
This paper uses official data from metropolitan regions to make a quantitative and dynamic exercise that simulates three economic markets, imposes taxes on them, cumulatively validates the model, and tests alternatives for fiscal distributions among municipalities.

Thus, this text provides additional quantitative arguments to guide the efficiency and effectiveness within metropolitan areas for the Brazilian case. In fact, based on the economic concept of metropolitan functional regions, the literature reports an unequal distribution, to the detriment of the peripheries, in relation to violence, congestion, access to public services, and restriction of opportunities. The understanding that such inequality affects the quality of life of the municipalities motivated us to investigate alternatives of tax distribution. 

The analysis described in this paper allows us to suggest two central conclusions. The first is that the progressiveness of MPF is striking in the metropolitan regions and its maintenance is significantly positive. The second conclusion is that in most metropolitan regions municipal merger for tax purposes would be beneficial. The effect, however, is not homogeneous and would have to be verified on a case-by-case basis. In some metropolitan regions, the gain of the merger is so relevant that it would be sufficient to compensate for the hypothesis of absence of the MPF.

In feasible terms, Constitutional Amendment number 15 of 1996 supports the possibility of municipal mergers, subject to consultation with the populations concerned. 

Finally, it should be noted that federalism with a municipal emphasis given by the Federal Constitution of 1988 may be advantageous for many Brazilian municipalities, spread over a continental country. Its excessive fragmentation, in the metropolitan scope in particular, however, seems to be more detrimental than beneficial. We hope that the indicatives presented in this paper contribute to the accumulation of evidence for better management and governance of urban space and its economic implications for the Brazilian case. 


\bibliographystyle{acm}
\end{document}